\documentclass[11pt]{article}
\usepackage{graphicx}

\newcommand{\BABARPubYear}    {06}
\newcommand{\BABARConfNumber} {001}
\newcommand{\SLACPubNumber} {12026}

\input babarsym

\long\def\inst#1{\par\nobreak\kern 4pt\nobreak
    {\it #1}\par\vskip 10pt plus 3pt minus 3pt}

\begin{document}
{\pagestyle{empty}

\begin{flushright}
\babar-CONF-\BABARPubYear/\BABARConfNumber \\
SLAC-PUB-\SLACPubNumber \\
\end{flushright}

\par\vskip 5cm

\begin{center}
\Large \bf 
A search for the decays {\boldmath $\Bp\to e^+ \nu_e$} and {\boldmath $\Bp\to \mu^+ \nu_\mu$} using
hadronic-tag reconstruction
\end{center}
\bigskip

\begin{center}
\large The \babar\ Collaboration\\
\mbox{ }\\
\today
\end{center}
\bigskip \bigskip

\begin{center}
\large \bf Abstract
\end{center}
We report on a search for the rare decay modes $\Bp\to e^+ \nu_e$ and $\Bp\to \mu^+ \nu_\mu$
with data collected from the \babar\ detector
at the PEP-II \epem\ storage ring. This search
utilizes a new technique in which we fully reconstruct the accompanying
$\Bm$ in \FourS\to\BpBm events, and look for a mono-energetic lepton in the \Bp rest frame.
No signal candidates are observed in either of the channels, consistent with the expected background, 
in a data sample of approximately 229 million \BB pairs. The branching-fraction upper limits are set
at $\mathcal{B}(\Bp\to\ep\nue)<7.9 \times 10^{-6}$ and $\mathcal{B}(\Bp\to\mup\num)<6.2 \times 10^{-6}$
at the 90\% confidence level.
\vfill
\begin{center}

Submitted to the 33$^{\rm rd}$ International Conference on High-Energy Physics, ICHEP 06,\\
26 July---2 August 2006, Moscow, Russia.

\end{center}

\vspace{1.0cm}
\begin{center}
{\em Stanford Linear Accelerator Center, Stanford University, 
Stanford, CA 94309} \\ \vspace{0.1cm}\hrule\vspace{0.1cm}
Work supported in part by Department of Energy contract DE-AC03-76SF00515.
\end{center}

\newpage
} 

\begin{center}
\small

The \babar\ Collaboration,
\bigskip

%
{B.~Aubert,}
{R.~Barate,}
{M.~Bona,}
{D.~Boutigny,}
{F.~Couderc,}
{Y.~Karyotakis,}
{J.~P.~Lees,}
{V.~Poireau,}
{V.~Tisserand,}
{A.~Zghiche}
\inst{Laboratoire de Physique des Particules, IN2P3/CNRS et Universit\'e de Savoie,
 F-74941 Annecy-Le-Vieux, France }
% ...

\end{center}\newpage

\section{INTRODUCTION}
\label{sec:Introduction}
Leptonic decays of \B mesons are of interest both for their use in characterizing Standard Model (SM) processes
and as probes of possible physics beyond the SM. They are, however, experimentally challenging to observe
due to both their small branching fractions and the difficulty of measuring decay modes with 
neutrinos in the final state. In the SM, these leptonic decays may proceed via tree-level processes mediated by
a virtual \Wp boson. The SM branching fraction for this type of decay\cite{charge} is given by
\begin{equation}
        \mathcal{B}(\B^+\rightarrow l^+\nu_l) =
        \frac{G_F^2 m_B m_l^2}{8\pi}\left(1-\frac{m_l^2}{m_B^2}\right)
        f_B^2 |V_{ub}|^2 \tau_B\textrm{,}
\label{eq:SMBranch}
\end{equation}
where $G_F$ is the Fermi coupling constant, $m_l$ and $m_B$ are the lepton and
\B-meson masses, and $\tau_B$ is the \B lifetime. Measurements of the leptonic branching fractions can
therefore constrain the product of \Vub and $f_B$, where \Vub is the Cabibbo-Kobayashi-Maskawa
matrix element parameterizing the $b \to u$ coupling and $f_B$ is a decay constant describing the wavefunction
overlap of the two quarks of the \B meson. \Vub has been obtained from analyses of semileptonic B decay
modes \cite{BBmixing}.  A measurement of the branching fractions in Eq.~\ref{eq:SMBranch} would allow us to
experimentally constrain $f_B$, which can be calculated in lattice QCD with
uncertainties of $\sim$15\%~\cite{lattice}. Measuring a product $\Vub \times f_B$, that is
inconsistent with measurements using other methods, could be interpreted as evidence for new physics. An enhancement
to the branching fraction could be caused by the existence of additional contributions,
for example as predicted by supersymmetry \cite{supersym}. 

As shown in Eq.~\ref{eq:SMBranch}, the decay rates are helicity suppressed
by a factor of $m_l^2/m_B^2$, yielding SM branching fraction predictions
of $\order (10^{-12})$ and $\order (10^{-7})$ for the $\electron$ and $\mu$ modes respectively.
The most stringent upper limits on the leptonic decay modes \Bp\to\ep\nue and \Bp\to\mup\num
are $1.5\times10^{-5}$ \cite{enu} and $6.6\times10^{-6}$ \cite{munu}, respectively,
at the 90\% confidence level.  A preliminary results of $\mathcal{B}(\Bp\to\ep\nue)<5.4\times10^{-6}$ and
$\mathcal{B}(\Bp\to\mup\num)<2.0\times10^{-6}$ at the 90\% confidence level \cite{prelims} 
are also available from the Belle Collaboration.

We search for the decays $\Bp\to e^+ \nu_e$ and $\Bp\to \mu^+ \nu_\mu$
using a technique in which the accompanying \Bm is reconstructed exclusively
in one of several hadronic decay modes. This technique has previously been applied to other
rare-decay searches~\cite{taunu, Knunu}, but it has not been used previously in a
search for $\Bp\to\ellp\nul$. Although the hadronic-reconstruction procedure has a relatively low efficiency,
this method has the advantages of highly suppressed backgrounds and
knowledge of the signal-lepton energy. Because it is very unlikely to observe
any background events, an observation of events in the signal region would be highly suggestive of a signal.

The leptonic decay $\Bp\to\taup\nut$ is expected to have a branching
fraction of $\order (10^{-4})$ \cite{UTfit}. Although this decay is less suppressed than the $e$ and $\mu$ modes,
analysis of it is more difficult due to additional neutrinos in the final state.
The preliminary result from Belle Collaboration (using a similar tagged-\B method) reports first evidence
for $\Bp\to\taup\nut$, with a branching fraction of
$1.06^{+0.34}_{-0.28}$ (stat) $ ^{+0.18}_{-0.16}$ (syst) $\times 10^{-4}$ \cite{Belletau}. As the $e$ and $\mu$ modes
are experimentally cleaner, they may ultimately yield more accurate branching-fraction results than $\Bp\to\taup\nut$
even with the higher suppression.

\section{THE \babar\ DETECTOR AND DATASET}
\label{sec:babar}
The on-resonance data sample used in this analysis corresponds to an integrated luminosity of
$208.7\,\mbox{fb}^{-1}$, accumulated at the \FourS resonance.
The events were recorded by the \babar\ detector at the \pep2\ asymmetric \epem\ storage ring.
In addition to the on-resonance sample, an off-resonance sample of $21.5\,\mbox{fb}^{-1}$
was recorded approximately $40\,\mbox{MeV}$ below the \FourS resonance, which is used for studies of
backgrounds not originating from the decay of \FourS.

The \babar\ detector is described in greater detail in the literature~\cite{babar}.
Charged-particle tracking and $dE/dx$ measurements for particle identification are provided by both a five-layer
double-sided silicon vertex tracker and a 40-layer drift chamber contained within the magnetic field of a
$1.5\,\mbox{T}$ superconducting solenoid. A ring-imaging Cherenkov detector (DIRC) provides efficient particle 
identification.
The energies of neutral particles are measured by an electromagnetic calorimeter consisting of 6580 CsI(Tl)
crystals. Muon identification is provided by resistive plate chambers.
Signal efficiencies and background rates are estimated using a Monte Carlo (MC) simulation of the \babar\ detector
based on GEANT4 \cite{geant}.

\section{ANALYSIS METHOD}
\label{sec:Analysis}
\begin{figure}[!htb]
\begin{center}
\includegraphics[height=7cm]{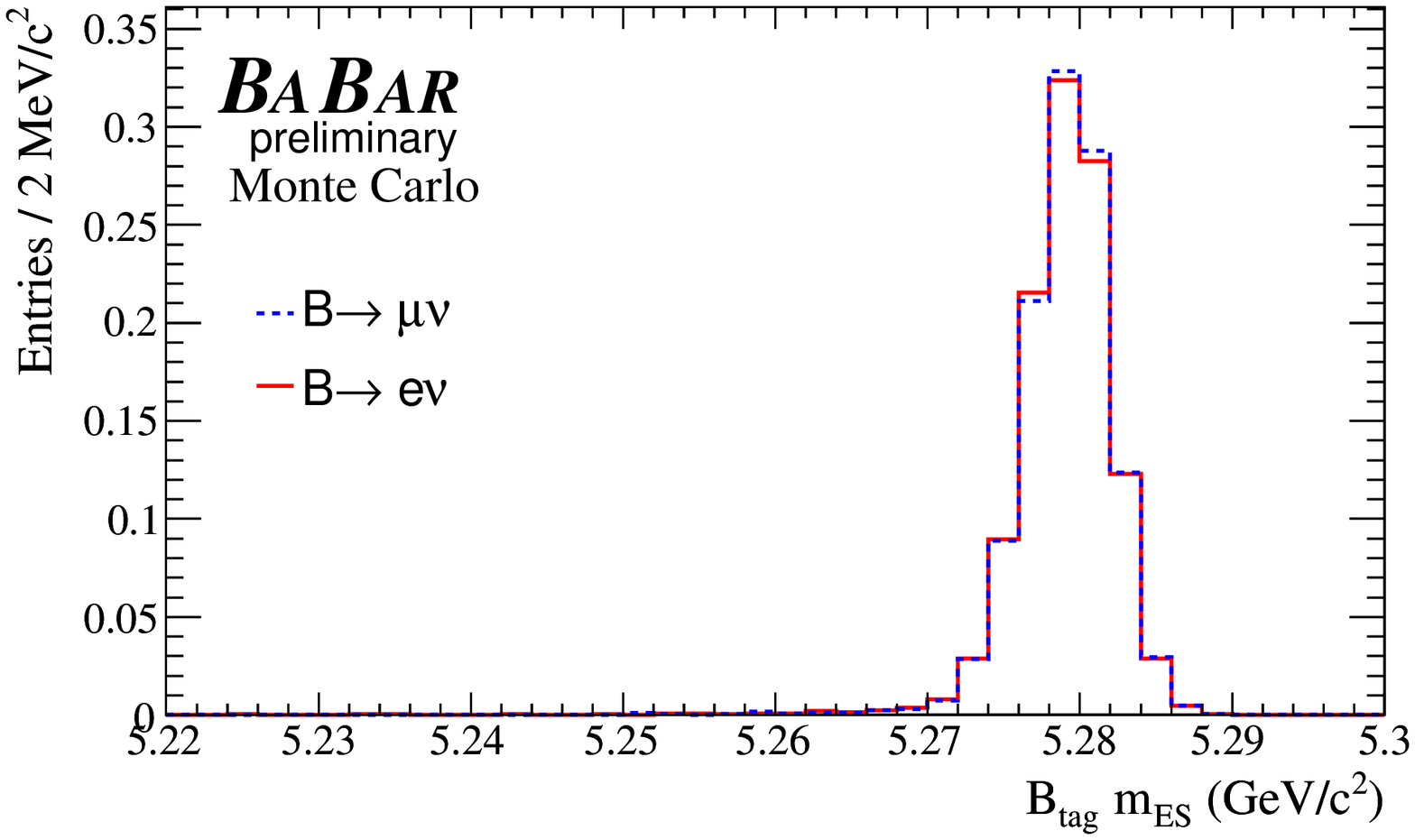}
\includegraphics[height=7cm]{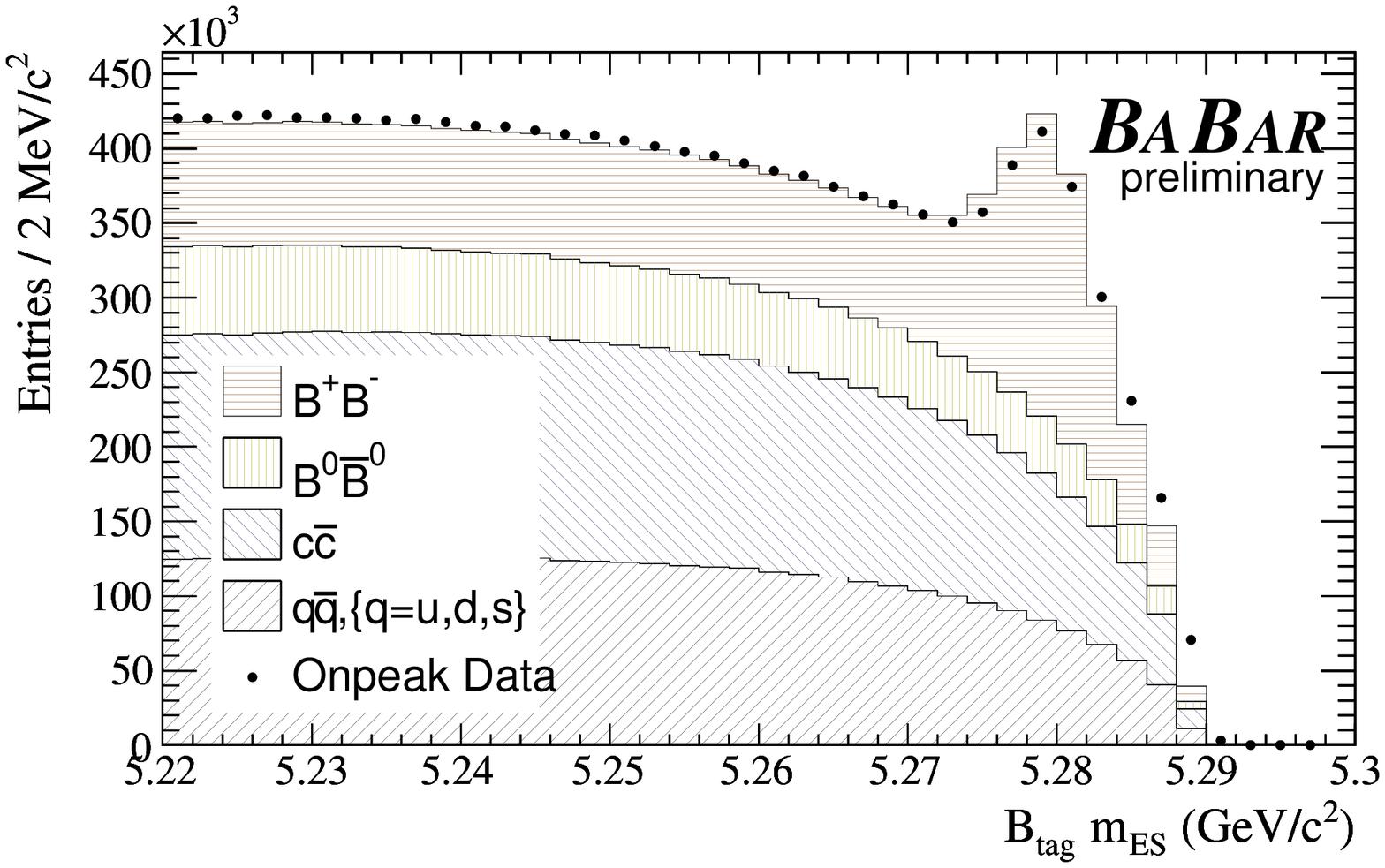}
\caption{$m_{ES}$ distribution of $B_{\rm tag}$ candidates. Signal MC events are shown on
top and background MC events on the bottom. The events are required to have
one reconstructed charged \B. The signal and background samples are scaled to data luminosity,
assuming both the signal mode branching fractions to be $2.0\times10^{-6}$
(\textit{i.e.}, the current upper limit of the mode $\Bp\to\mup\num$).}
\label{fig:mes}
\end{center}
\end{figure}

The decay $B^+\to \ell^+ \nu_\ell$ produces a mono-energetic charged $e$ or $\mu$ in the \Bp rest frame,
accompanied by the missing-energy signature of a neutrino. To gain sensitivity to this signature,
a \B meson from the decay $\Upsilon(4S)\rightarrow\BB$, $\B_{\rm tag}$,
is reconstructed in hadronic decay modes \Bp\to$\bar{D}^{(*)0}X^{+}$, \Bm\to$D^{(*)0}X^{-}$, 
\Bz\to$D^{(*)-}X^{+}$ and $\bar{B}^{0}$\to$D^{(*)+}X^{-}$ \cite{semiexcl}. While both charged and neutral $\B_{\rm tag}$ candidates are considered 
for reconstruction, the events where the best $\B_{\rm tag}$ candidate (as discussed below) is neutral are vetoed.
\Dz candidates are reconstructed in the modes $\Dz\to\Km\pip$, 
$\Dz\to\Km\pip\piz$, $\Dz\to\Km\pip\pip\pim$ and $\Dz\to\KS\pip\pim$; \Dstarz candidates are 
formed from a \Dz and either a \piz or a $\gamma$. Similar reconstruction is done for the $D^{(*)\pm}$
candidate. Pions and kaons remaining in the event are combined in subsets with the $D$ candidate to form
a $\B_{\rm tag}$ candidate, considering the quantity $|\Delta E| = |E_B - E_{\rm beam}|$, where
$E_B$ is the energy of the reconstructed \B and $E_{\rm beam}$ is the beam energy, both in the center-of-mass 
(CM) frame. $\B_{\rm tag}$ candidates must satisfy the requirement $|\Delta E|<0.2\,\mbox{GeV}$,
and the $\B_{\rm tag}$ with the lowest value of $|\Delta E|$ is chosen as the candidate.
In the case of multiple $\B_{\rm tag}$ candidates, the one with a highest {\em a priori} purity is chosen. 
The energy-substituted mass
\begin{equation}\label{eq:mES1}
        m_{ES} = \sqrt{E_{\rm beam}^2 - {\vec{p}_B}^2},
\end{equation}
where $\vec{p}_B$ is the momentum of the \B candidate in the CM frame,
is required to be
$5.270\,\mbox{GeV}/c^2<m_{ES}<5.288\,\mbox{GeV}/c^2$. The $m_{ES}$
distribution is shown in Fig.~\ref{fig:mes}. The resolution in $m_{ES}$ is dominated by the spread in
$E_{\rm beam}$, $\sigma_{E_{\rm beam}} \approx 2.59\,\mbox{MeV}$. 

Event shape requirements are used to suppress the combinatorial backgrounds not originating
from the decay of \FourS. These requirements are: $R2<0.5$, where $R2$ is the ratio of zeroth and the
second Fox-Wolfram moment \cite{foxwolf}; $|\cos\theta_T|<0.9$, where $\theta_T$ represents
the angle between the thrust axis defined by the $\B_{\rm tag}$ candidate and by the combination of all
other tracks and clusters in the event. 

\begin{figure}[!htb]
\begin{center}
\includegraphics[height=7cm]{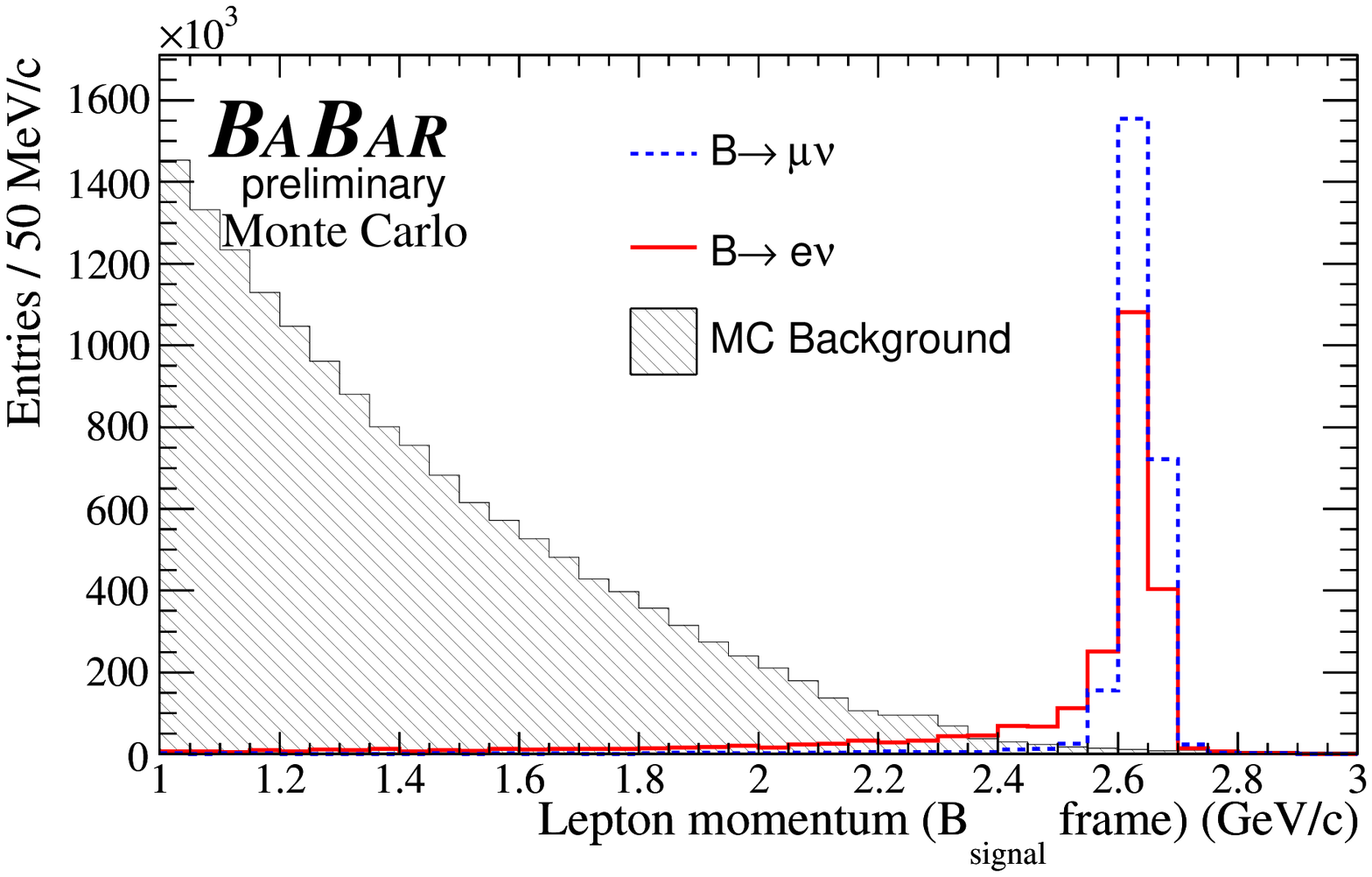}
\includegraphics[height=7cm]{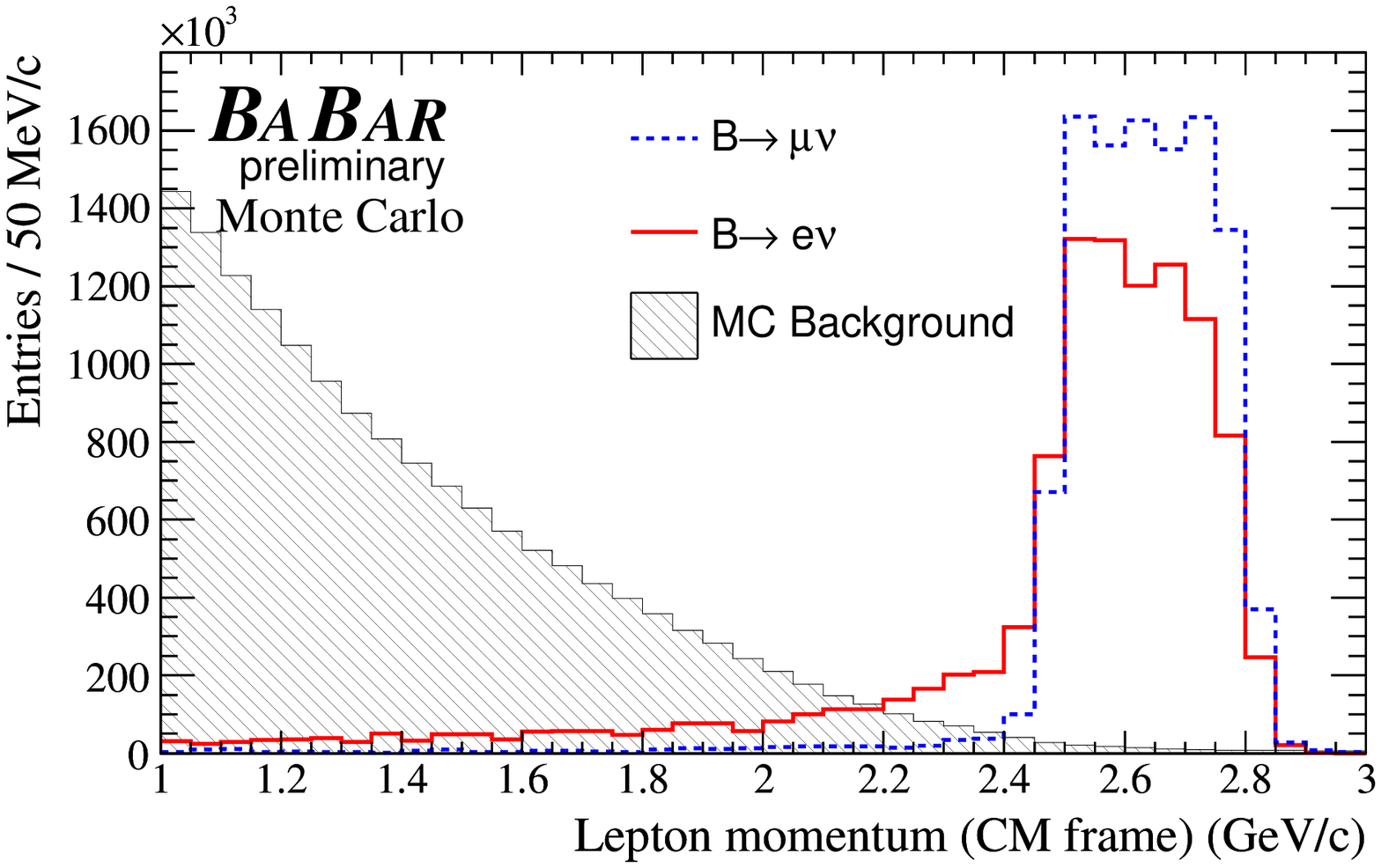}
\caption{The lepton-candidate momentum distributions in the $B_{\rm signal}$
frame (top) and in the CM frame (bottom) before applying the signal-selection criteria.
The resolution gain provided by the $\B_{\rm tag}$ reconstruction allows a tighter
selection around the signal peak. The signal MC is scaled arbitrarily.}
\label{fig:pcompare}
\end{center}
\end{figure}

The tracks and clusters not used in the  $\B_{\rm tag}$ reconstruction are assumed to originate
from the decay of the signal \B, $\B_{\rm signal}$. Since the CM energy is precisely known, reconstruction of
the $\B_{\rm tag}$ fully determines the energy and momentum, and thus the rest frame, of the $\B_{\rm signal}$ 
candidate. The signal lepton energy is constrained to be at the kinematic endpoint in the $B_{\rm signal}$ rest frame,
producing a distinct experimental signature. Fig.~\ref{fig:pcompare}
shows a comparison of the lepton-momentum distributions in the $\B_{\rm signal}$ rest frame and the CM frame.
The $\B_{\rm signal}$ rest frame is calculated from the recorded beam energies, and
a momentum opposite in direction to the reconstructed $\B_{\rm tag}$.
We require the highest momentum track of correct charge in the $\B_{\rm signal}$ frame to have a momentum $\vec{p^*}$ between
2.54 and 2.72 GeV/$c^2$. This highest momentum track
is also required to satisfy ($e$ or $\mu$) lepton identification requirements, and to survive a veto on kaon 
particle identification based primarity on DIRC.

The missing momentum in the $\B_{\rm signal}$ frame is measured to be
\begin{equation}\label{eq:pmissdef}
        \vec{p^{*}}_{\rm miss} = \vec{p^{*}}_{\FourS} - \vec{p^*}_{B_{\rm tag}} - \sum_i {\vec{p^*}_{i}}\textrm{,}
\end{equation}
where the sum is of all charged and neutral particles not associated with the reconstruction of the
$\B_{\rm tag}$, and all quantities are computed in the $\B_{\rm signal}$ rest frame. The direction of 
the missing momentum is described by
\begin{equation}\label{eq:pmiss}
        \cos{\theta_{p}}_{\rm miss} = \frac{p_{z_{\rm miss}}}{|\vec{p}_{\rm miss}|}\textrm{,}
\end{equation}
where $p_{z_{\rm miss}}$ is the component of the momentum relative to the axis defined by the $e^{-}$ (high energy) beam 
direction, and the momenta are 
computed in the CM frame. To ensure that the missing momentum is carried
by the neutrino rather than a particle passing outside of the detector acceptance,
we require $-0.76<\cos\theta_{p_{\rm miss}}<0.92$.
			
We also require that the missing momentum be consistent with
a neutrino recoiling against the signal candidate lepton by
requiring $\Delta P^{*}_{\rm miss}<0.7\,\mbox{GeV/c}$, where
\begin{equation}\label{eq:dP}
        \Delta P^{*}_{\rm miss} = | \vec{p}^{*}_{\rm miss} + \vec{p^*} |\textrm{.}
	\end{equation}
This requirement is useful in suppressing semileptonic \B decays,
where a high momentum lepton may be present.

The amount of extra energy in the event is described by a quantity $E_{\rm extra}$, defined by
\begin{equation}\label{eq:Eextra}
        E_{\rm extra} = \sum_i E_i\textrm{,}
\end{equation}
where $E_{i}$ is the CM frame energy of a given particle, and the sum runs over all tracks and clusters not associated 
with the $\B_{\rm tag}$ or the lepton candidate. We select events with $E_{\rm extra}<1.2\,\mbox{GeV}$.
The resulting selection efficiencies after applying the above selection criteria are $(0.122\pm0.012)\,\%$
and $(0.145\pm0.013)\,\%$ for $e$ and $\mu$ modes, respectively, where the uncertainties
include only MC statistics.

Backgrounds potentially arise either from sources with a peak in $\B_{\rm tag}$ $m_{ES}$, or which are combinatorial
in $\B_{\rm tag}$ $m_{ES}$. One source of the nonpeaking background is from the misreconstruction of the 
decay products of \tautau or $q\bar{q}$ ($q=u,d,s,c$). These backgrounds are strongly suppressed by event
shape requirements. Nonpeaking background can also arise from misreconstructed $\B_{tag}$ candidates in \BB events. 

No nonpeaking background events in MC (\tautau and $q\bar{q}$ samples are approximately equal to the data 
luminosity, and \BB sample is $\sim5$ times the data sample) pass the selection criteria. The absence of 
nonpeaking background is verified by studying the $p^*$ and $m_{ES}$ sidebands 
($2.0\,\mbox{GeV}<p^*<2.5\,\mbox{GeV}$ and $5.22\,\mbox{GeV}/c^2<m_{ES}<5.26\,\mbox{GeV}/c^2$).
The ratios of the number of events in these sidebands are used to extrapolate the background level in the signal region.
For \BpBm events, the nonpeaking contribution is estimated by fitting the $m_{ES}$ distribution with a
combination of a Gaussian and an ARGUS function \cite{argus}. The background estimate from the sidebands is
consistent with the prediction from the MC of no nonpeaking background.
As can be seen in Fig.~\ref{fig:mes}, the $m_{ES}$ sideband region is well
modeled by background MC up to the uncertainties arising from the relative cross sections of various decay modes.
For nonpeaking events, the quantities $p^*$ and $m_{ES}$ are assumed to be uncorrelated in both the signal and
sideband regions.

In peaking background events, $\B_{\rm tag}$ is correctly reconstructed, but the $\B_{\rm signal}$ decays via 
a non-signal mode. The most likely backgrounds are $b\to u\ell\nul$ events and two-body decays with a misidentified 
$\pi$, where the signal lepton may be near the kinematic endpoint. The total \BpBm background (nonpeaking 
and peaking together) is estimated by counting the number of events in the MC sample
that pass the selection criteria. Requirements on the signal-lepton momentum and the missing momentum
are especially important in discerning the signal events.
After rescaling as discussed below, the \BpBm contribution to the
background is estimated to be $0.115^{+0.131}_{-0.075}$ and $0.229^{+0.167}_{-0.142}$ events per $209\,\mbox{fb}^{-1}$,
for $e$ and $\mu$ modes respectively. Since the contributions from the other background types are found to be
negligible, the above quantities are taken as the total background estimates.

\section{SYSTEMATIC STUDIES}
\label{sec:Systematics}
The study of the lepton-momentum sidebands reveals a slight discrepancy between the $m_{ES}$ peaking
\BpBm yields in data and in MC. To estimate the $\B_{\rm tag}$ yield in the samples, the $m_{ES}$ distribution
for the on-resonance data sample is studied after subtracting the normalized off-resonance $m_{ES}$ 
distribution from it. To compensate for the CM energy difference, the off-resonance
$m_{ES}$ distribution is shifted by $\sim20\,\mbox{MeV}$, matching the kinematic endpoints of
the two distributions. The precise values of the required energy shift are determined by comparing the
on-resonance and off-resonance CM energy distributions.
By fitting a sum of a Gaussian and an ARGUS function to the background subtracted $m_{ES}$ distribution,
we determine the parameters for the shape of the nonpeaking contribution.
Another set of such parameters is estimated by a similar fit using a simulated \BB sample.
These parameters are used to fit the signal region in the data and MC \BB samples.
The yields are calculated by integrating the Gaussian component of the resulting
fits, resulting in a range of values for the yield (see Table~\ref{tab:yields}). The scaling
factor, which is the ratio of the data and the MC yields, is calculated to be $0.908\pm0.013(stat)\pm0.015(syst)$,
where the systematic error arises from the uncertainties involved in the fitting procedure. It is notable that
while the yield itself has a large uncertainty of $\sim8.8\,\%$, the ratio between the data and MC
yields are less dependent on the exact shape of the nonpeaking contribution.
Applying the correction factor to quantities determined from the \BB MC sample 
introduces a systematic error. However, the signal efficiency and the background estimates determined 
from MC samples are statistically limited.

\begin{table}[htbp]
\caption{The peaking \BB yields resulting from using different ARGUS function shapes for
describing the combinatorial contribution. The yields are expressed in \BpBm pairs per $\mbox{fb}^{-1}$.}
\begin{center}
\begin{tabular}{|c|c|c|c|} \hline
ARGUS shape source& MC yield & Data yield & correction  \\ \hline
From Data               & 2253.38          & 2018.25            & 0.896 \\
From MC sample          & 2581.89          & 2380.38            & 0.922 \\ \hline
Average                 &                  &                    & 0.908$\pm$0.013 \\ \hline
\end{tabular}
\end{center}
\label{tab:yields}
\end{table}

Tracking efficiency uncertainties are considered, introducing an additional $0.8\,\%$/track
systematic error on the signal efficiency and the background estimate. The lepton-identification efficiencies
for data and MC are compared and correction factors of $0.96\pm0.01$ and $0.962\pm0.015$ are
determined for $e$ and $\mu$ respectively. This correction factor affects all
quantities determined from MC samples, introducing a systematic uncertainty.
The misidentification rate of leptons as pions is also studied. The difference between the misidentification rates
between data and MC introduces a correction factor to background estimates. In the case of muons,
the pion misidentification rate is estimated to be $(5.0\pm0.5)\,\%$ for MC and $(5\pm1)\,\%$ for data. 
For the electron identification,
the discrepancy is larger with misidentification rates of $(0.01\pm0.01)\,\%$ in MC and $(0.05\pm0.02)\,\%$ in data.
This difference requires the number of background events, estimated from MC, to be scaled appropriately. However,
as all MC events that pass the selection criteria are verified to have a correct lepton identification,
we do not include a misidentification-rate correction to the background estimate. Additional systematic errors,
associated with the quantities $E_{\rm extra}$ and $\Delta P^{*}_{\rm miss}$, are estimated by varying the selection 
criteria by $100\mev$. The variations in the signal efficiencies are found to be very small, requiring no further 
additions to the systematics.

The efficiencies and the background estimates, after applying all corrections,
are listed in Table~\ref{tab:listofnumbers}. The $\epsilon_{\rm tag}$ and $\epsilon_{\rm sig}$ correspond
to the efficiencies of the $\B_{\rm tag}$ reconstruction in the signal region and the
signal-lepton selection respectively. The $\epsilon_{\rm tot}$ represents the product of these
two quantities. The quantities $N_{\rm bg}$, $N_{\rm SM}$ and $N_{\rm obs}$ represent, respectively, the 
number of estimated background events, the expected number of signal events according 
to the SM branching fractions and the number of signal candidates observed in the data.

\begin{table}[htbp]
\caption{Efficiencies and background estimates, used for calculating the branching fractions, after applying all corrections.
The first error is statistical, the second is systematic.}
\begin{center}
\begin{tabular}{|c|c|c|} \hline
Quantity       & \B\to\mup\num  &  \B\to\ep\nue  \\   \hline
$\sigma_{\BB}$ & \multicolumn{2}{|c|}{1.05nb} \\  \hline
$\mathcal{L}$  & \multicolumn{2}{|c|}{208.7$fb^{-1}$} \\ \hline
$N_{\BB}$   & \multicolumn{2}{|c|}{229.953$\times 10^6$} \\ \hline
$\epsilon_{\rm tag}$ $(\%)$ & $0.239\pm0.013\pm0.004$ & $0.247\pm0.013\pm0.004$ \\
$\epsilon_{\rm sig}$ $(\%)$ & $60.5\pm4.0\pm1.0$  & $49.4\pm3.8\pm0.8$  \\
$\epsilon_{\rm tot}$ $(\%)$ & $0.145\pm0.013\pm0.003$ & $0.122\pm0.012\pm0.003$  \\
$N_{\rm bg}$         & $0.229^{+0.167}_{-0.142}\pm0.007$ & $0.115^{+0.131}_{-0.075}\pm0.004$ \\
$N_{\rm SM}$         & $\sim 0.03$                       & $\sim 3\times 10^{-7}$ \\
$N_{\rm obs}$        & 0 & 0 \\ \hline
\end{tabular}
\end{center}
\label{tab:listofnumbers}
\end{table}

\section{PHYSICS RESULTS}
\label{sec:Physics}
\begin{figure}[!htb]
\begin{center}
\includegraphics[height=7cm]{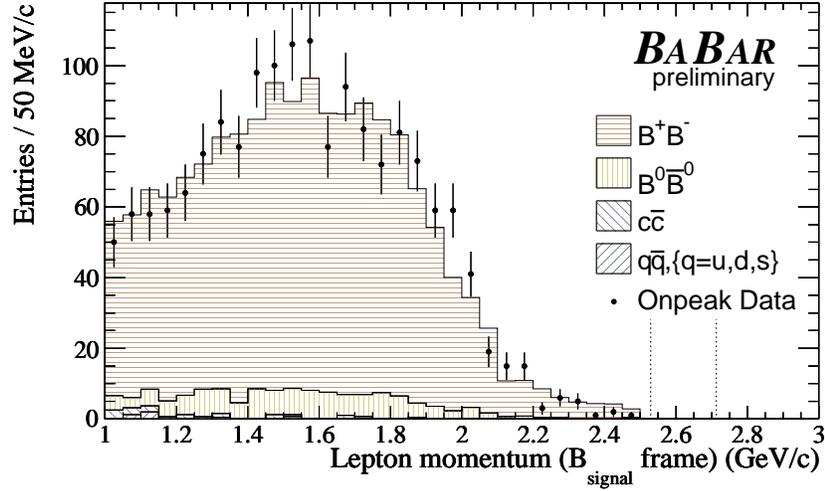}
\caption{Signal-lepton momentum in $B_{\rm signal}$ frame after all signal
selection criteria. No events in data are present in the signal region.}
\label{fig:finalpstar}
\end{center}
\end{figure}

Fig.~\ref{fig:finalpstar} shows the momentum distribution of the signal lepton in the $\B_{signal}$ frame for data and
background MC. The absence of events in the signal region is consistent with SM expectations.

The branching fraction $\mathcal{B}$ is given by
\begin{equation}\label{eq:Branch}
        \mathcal{B}(B^+\rightarrow l^+\nu_l) =  \frac{N_{\rm obs}-N_{\rm bg}}{2\cdot N_{\BpBm}\cdot\epsilon_{tot}}\textrm{,}
	\end{equation}
where $N_{\rm obs}$ is the number of events that pass the selection, $N_{\rm bg}$ is the estimated background count,
$N_{\BpBm}$ is the number of \BpBm pairs in the data sample, and $\epsilon_{\rm tot}$ is
the total efficiency of the selection given by signal MC. The upper limit on $\mathcal{B}$ at the 90\%
confidence level is determined using a frequentist procedure that incorporates systematic uncertainties
in the signal efficiency and expected number of background events \cite{Barlow:2002bk}. This procedure yields
$\mathcal{B}(\Bp\to\ep\nue)<7.9 \times 10^{-6}$ and $\mathcal{B}(\Bp\to\mup\num)<6.2 \times 10^{-6}$
at the 90\% confidence level.
	
\section{SUMMARY}
\label{sec:Summary}
We have set the branching fraction upper limits for rare leptonic decays \Bp\to\ep\nue and \Bp\to\mup\num
to be $7.9 \times 10^{-6}$ and $6.2 \times 10^{-6}$ respectively, at the 90\% confidence level.
Both results are consistent with the Standard Model and represent improvements to the most stringent published
upper limits to date.

\section{ACKNOWLEDGMENTS}
\label{sec:Acknowledgments}

We are grateful for the 
extraordinary contributions of our \pep2\ colleagues in
achieving the excellent luminosity and machine conditions
that have made this work possible.
The success of this project also relies critically on the 
expertise and dedication of the computing organizations that 
support \babar.
The collaborating institutions wish to thank 
SLAC for its support and the kind hospitality extended to them. 
This work is supported by the
US Department of Energy
and National Science Foundation, the
Natural Sciences and Engineering Research Council (Canada),
Institute of High Energy Physics (China), the
Commissariat \`a l'Energie Atomique and
Institut National de Physique Nucl\'eaire et de Physique des Particules
(France), the
Bundesministerium f\"ur Bildung und Forschung and
Deutsche Forschungsgemeinschaft
(Germany), the
Istituto Nazionale di Fisica Nucleare (Italy),
the Foundation for Fundamental Research on Matter (The Netherlands),
the Research Council of Norway, the
Ministry of Science and Technology of the Russian Federation, and the
Particle Physics and Astronomy Research Council (United Kingdom). 
Individuals have received support from 
CONACyT (Mexico),
the A. P. Sloan Foundation, 
the Research Corporation,
and the Alexander von Humboldt Foundation.

\pagebreak

\appendix
\section{\boldmath Appendix}
Appended are additional plots of interest.

\begin{figure}[htbp]
\begin{center}
\includegraphics[height=7cm]{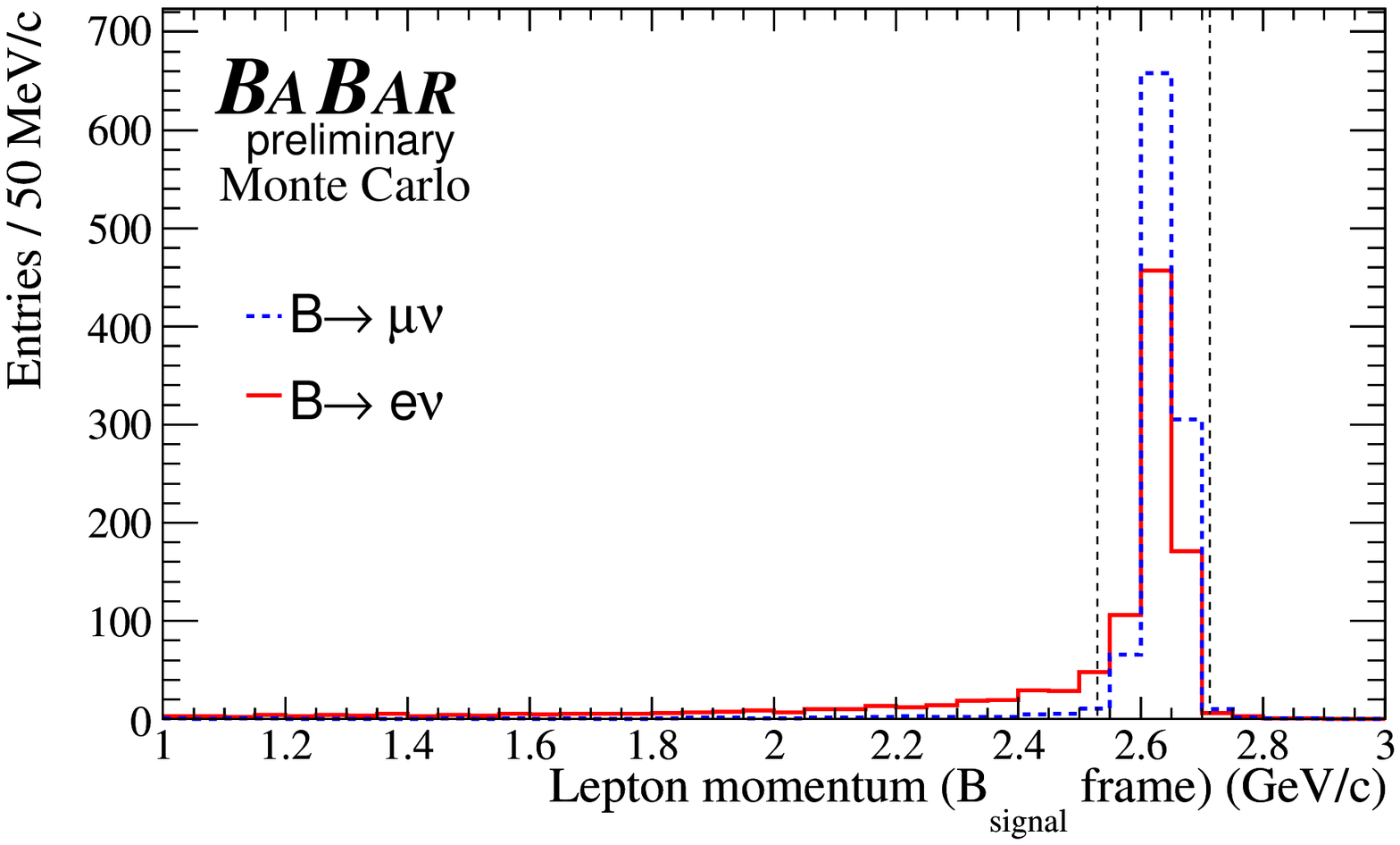}
\includegraphics[height=7cm]{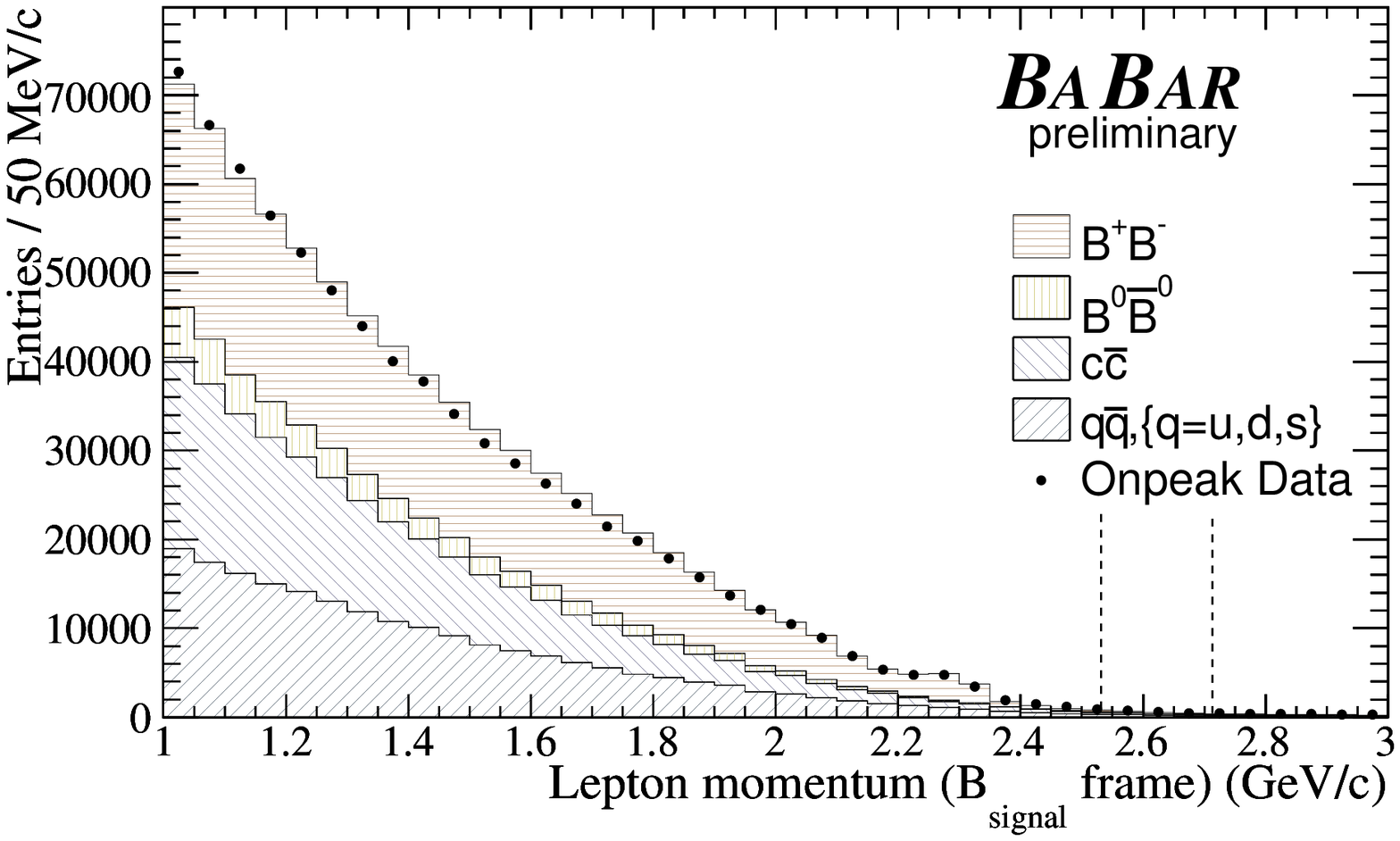}
\caption{The distribution of the lepton candidate momentum in the $B_{\rm signal}$
frame. The events are required to pass all reconstruction cuts.
The signal \Bp\to\mup\num and the 
backgrounds are scaled to onpeak data luminosity, assuming SM predictions for branching
fractions. The signal \Bp\to\ep\nue is scaled with respect to \Bp\to\mup\num
by the relative luminosity of the sample size.}
\label{fig:pstar}
\end{center}
\end{figure}

\begin{figure}[htbp]
\begin{center}
\includegraphics[height=7cm]{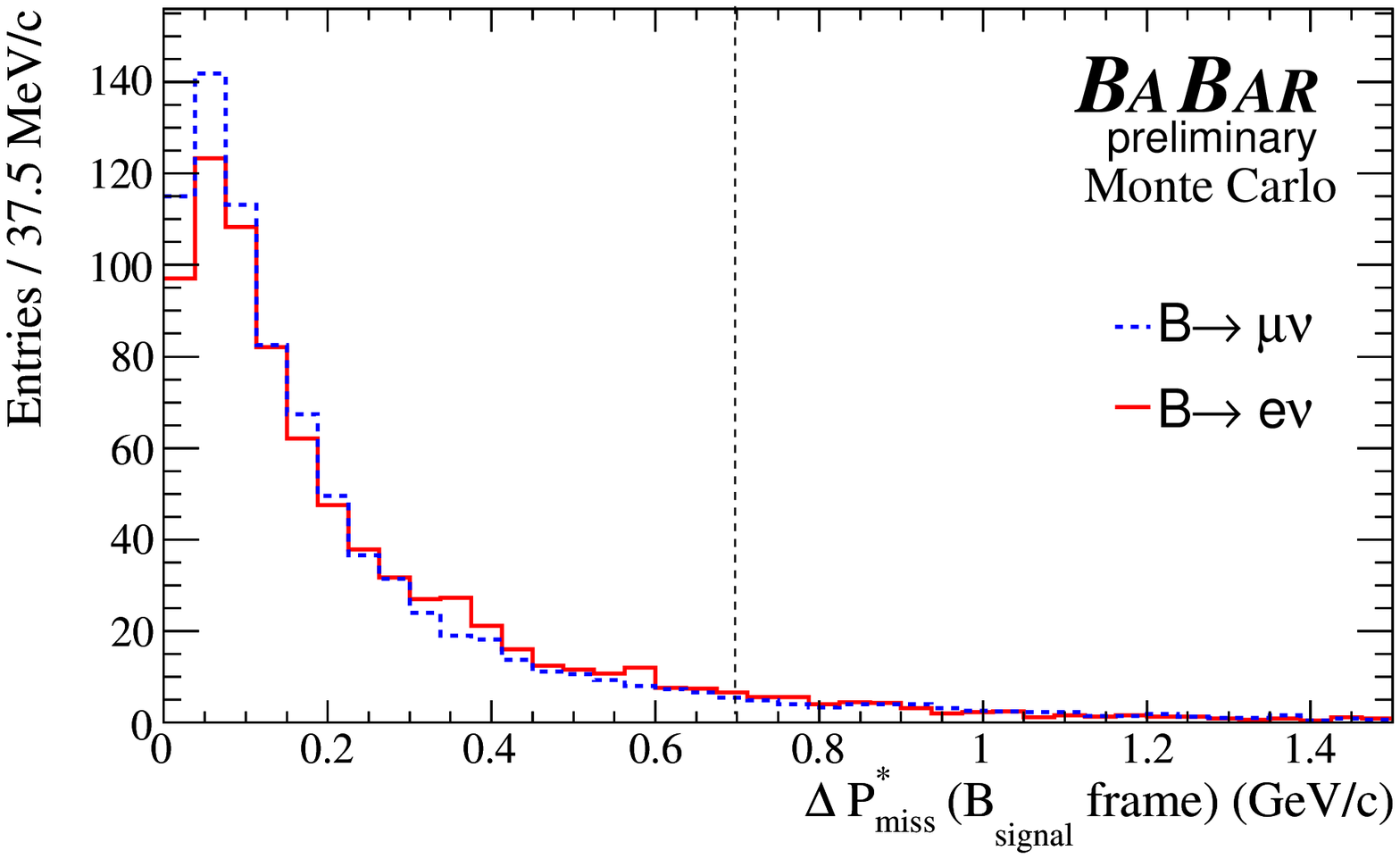}
\includegraphics[height=7cm]{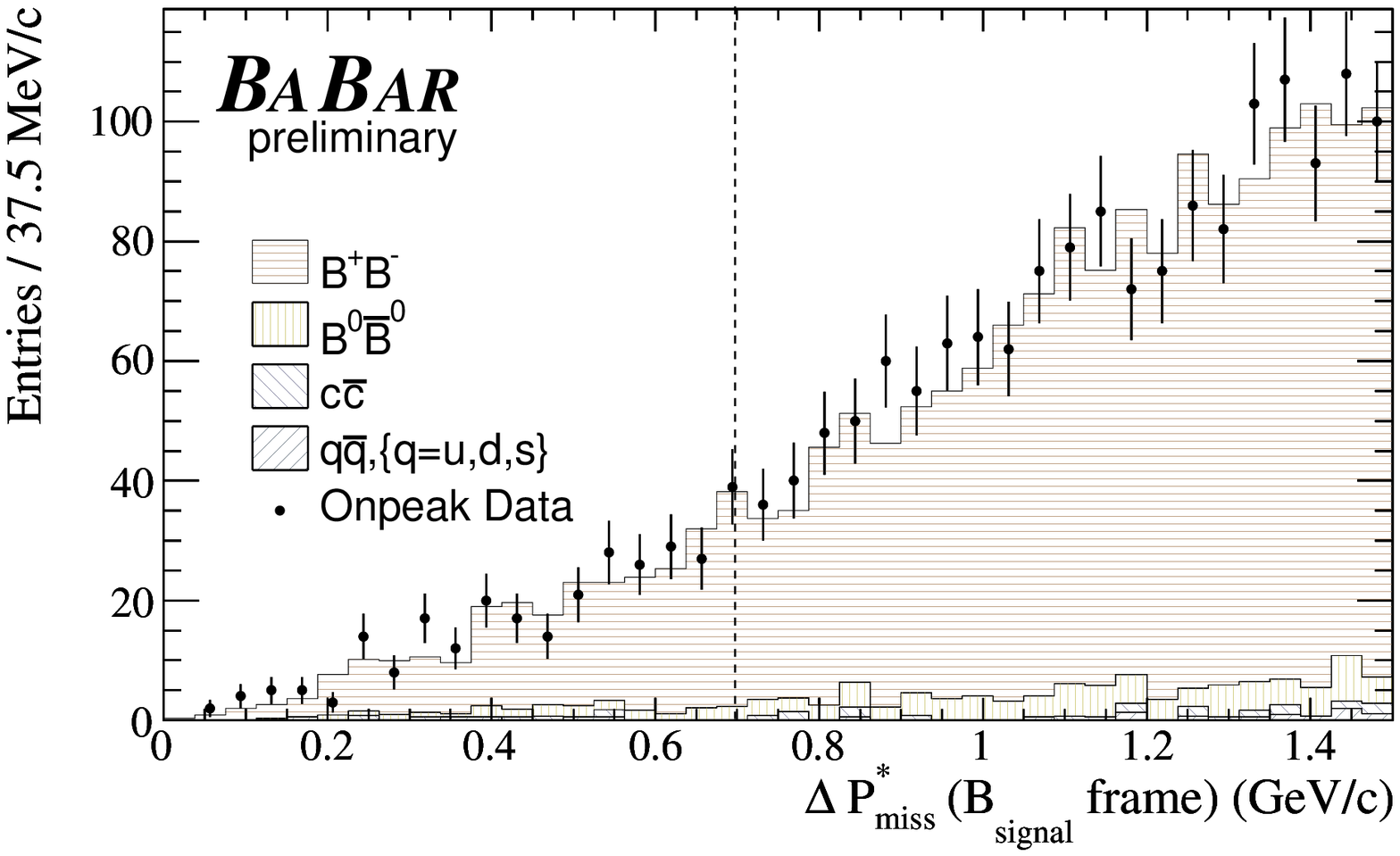}
\caption{The distribution of $\Delta P^{*}_{miss}$, for signal and background samples.
The events are required to pass all reconstruction and signal cuts (with a relaxed requirement of $p^*>2.0\gev$ to
increase background statistics). The signal \Bp\to\mup\num and the backgrounds are scaled to onpeak data,
assuming SM predictions for branching
fractions. The signal \Bp\to\ep\nue is scaled with respect to \Bp\to\mup\num
by the relative luminosity of the sample size.}
\label{fig:dP}
\end{center}
\end{figure}

\begin{figure}[htbp]
\begin{center}
\includegraphics[height=7cm]{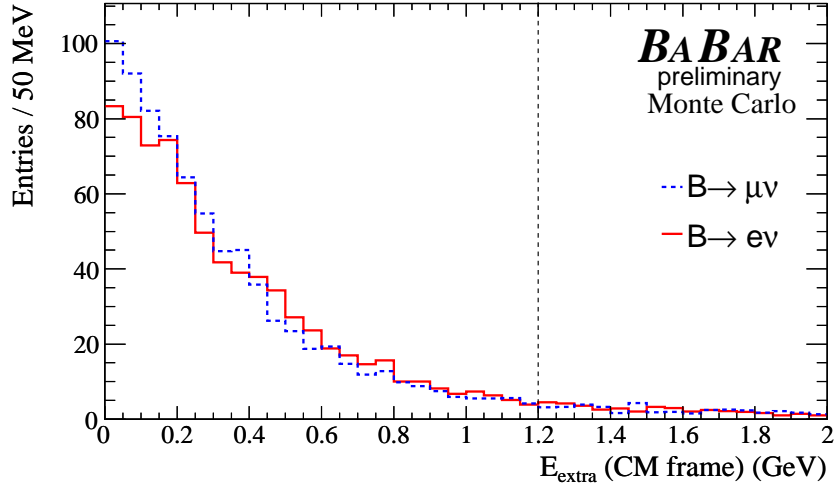}
\includegraphics[height=7cm]{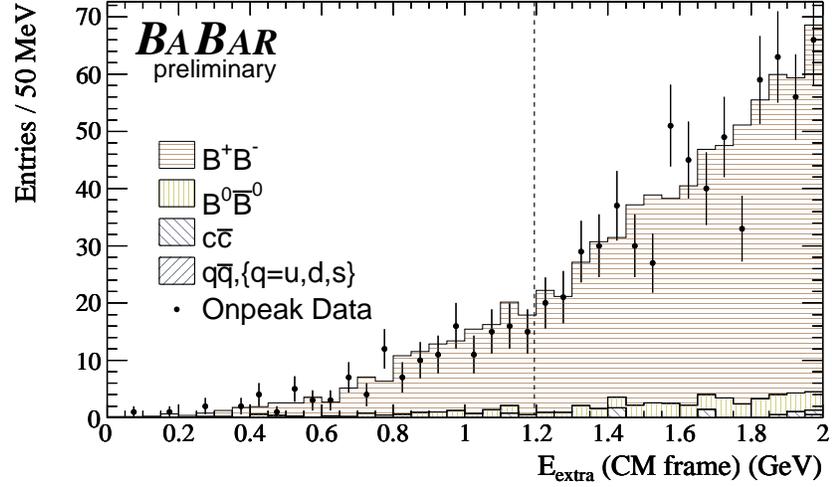}
\caption{$E_{\rm extra}$, the total energy not accounted for by the signal lepton
candidate, distribution. The events are required to pass all reconstruction
and signal cuts (with a relaxed requirement of $p^*>2.0\gev$ to
increase background statistics). The signal \Bp\to\mup\num and the
backgrounds are scaled to onpeak data luminosity, assuming SM predictions for branching
fractions. The signal \Bp\to\ep\nue is scaled with respect to \Bp\to\mup\num by the relative luminosity
of the sample size.}
\label{fig:Eextra}
\end{center}
\end{figure}

\end{document}